\documentclass[sigconf,nonacm]{acmart}

\AtBeginDocument{%
  \pagestyle{plain}%
  \flushbottom
  \brokenpenalty=100
  \setlength{\emergencystretch}{2em}%
  \setcounter{topnumber}{2}%
  \setcounter{bottomnumber}{1}%
  \setcounter{totalnumber}{3}%
  \setlength{\textfloatsep}{10pt plus 2pt minus 2pt}%
  \setlength{\floatsep}{8pt plus 2pt minus 2pt}%
  \setlength{\intextsep}{8pt plus 2pt minus 2pt}%
  \setlength{\dbltextfloatsep}{12pt plus 2pt minus 2pt}%
  \setlength{\dblfloatsep}{8pt plus 2pt minus 2pt}%
}

\AddToHook{env/thebibliography/begin}{%
  \AddToHookNext{shipout/after}{\balance}%
}

\usepackage{balance}
\usepackage{comment}
\usepackage{amsmath}
\usepackage{algorithm}
\usepackage{algpseudocode}
\usepackage{textcomp}
\usepackage{booktabs}
\usepackage{multirow}
\usepackage{makecell}
\usepackage{placeins}
\usepackage{float}
\usepackage{xspace}
\usepackage{siunitx}
\usepackage[normalem]{ulem}

\providecommand{\DIFadd}[1]{{\protect\color{blue}#1}}
\providecommand{\DIFdel}[1]{{\protect\color{red}\sout{#1}}}

\newcommand{\CompressionEdit}[2]{%
  \ifdefined\MAScopeBilingual
    \begingroup
      \color{red}%
      \renewcommand{\DIFadd}[1]{##1}%
      #1%
    \endgroup
    \DIFadd{#2}%
  \else
    #2%
  \fi
}

\providecommand{\CNtranslation}[2]{}

\providecommand{\BilingualDocumentBegin}{}
\providecommand{\BilingualDocumentEnd}{}

\newcommand{\sysname}{MAScope\xspace}
\newcommand{\TSE}{\textsf{TSE}\xspace}
\newcommand{\TCJudge}{\textsf{TC-Judge}\xspace}

\newcommand{\Layered}{Layered}
\newcommand{\Centralized}{Centralized}
\newcommand{\Decentralized}{Decentralized}
\newcommand{\SharedPool}{Shared Message Pool}

\newcommand{\SemanticDrift}{Semantic Drift}
\newcommand{\InfoFailure}{Information Loss}
\newcommand{\CollabStall}{Coordination Stall}
\newcommand{\VerifyMisalign}{Verification Misalignment}

\title[Know the Shape, Find the Fault]{Know the Shape, Find the Fault: Topology-Conditioned Diagnosis of Multi-Agent LLM Failures}

\author{Xinwen Liu}
\email{liuxw24@buaa.edu.cn}
\affiliation{
  \institution{Beihang University}
  \city{Beijing}
  \country{China}
}

\author{Zhuocheng Pan}
\email{zhuochengPan@buaa.edu.cn}
\affiliation{
  \institution{Beihang University}
  \city{Beijing}
  \country{China}
}

\author{Isabella Zhu}
\email{zhunini@buaa.edu.cn}
\affiliation{
  \institution{Beihang University}
  \city{Beijing}
  \country{China}
}

\author{Jiawei Zhang}
\email{jw\_zhang@buaa.edu.cn}
\affiliation{
  \institution{Beihang University}
  \city{Beijing}
  \country{China}
}

\author{Xudong Liu}
\email{liuxd@buaa.edu.cn}
\affiliation{
  \institution{Beihang University}
  \city{Beijing}
  \country{China}
}

\author{Tianyu Wo}
\email{woty@buaa.edu.cn}
\affiliation{
  \institution{Beihang University}
  \city{Beijing}
  \country{China}
}

\renewcommand{\shortauthors}{Liu Xinwen et al.}

\begin{document}
\BilingualDocumentBegin

\begin{abstract}
Multi-agent LLM systems coordinate task execution through exchanges
of information among agents. When coordination breaks down, similar
symptoms in execution traces can reflect different problems in how
information is passed, used, or verified. Communication topology
captures how agents exchange information and provides structural cues
for distinguishing coordination failure modes. Using these cues for
diagnosis requires establishing how topology relates to failure
patterns and recovering the relevant structure from execution traces
that lack explicit topology labels.
\CNtranslation{摘要一}{%
多智能体大语言模型系统通过智能体之间的信息交换协调任务执行。
当协作发生故障时，执行轨迹中相似的表象可能对应信息传递、使用或验证方式上的不同问题。
通信拓扑刻画智能体交换信息的方式，并为区分协作失败模式提供结构线索。
要将这些线索用于诊断，需要确定拓扑与失败模式之间的关系，
并从没有显式拓扑标签的执行轨迹中恢复相关结构。}

We analyze the relationship between communication topology and
failure patterns and introduce \sysname{}, a two-stage framework for
topology-conditioned diagnosis. Its Trace Structural Extractor
\TSE recovers communication topology from heterogeneous execution
traces by grounding an interaction graph in message evidence.
The Topology-Conditioned Judge \TCJudge then classifies failures
using the trace, predicted topology, an empirical failure prior
estimated from separate labeled traces, and a short description of
topology-specific failure patterns. Under a fixed orchestration
structure, the recovered topology can be reused across executions.
\CNtranslation{摘要二}{%
我们分析通信拓扑与失败模式之间的关系，并提出用于拓扑条件化诊断的两阶段框架 \sysname{}。
其中，轨迹结构提取器 \TSE 以消息证据为依据构建交互图，从异构执行轨迹中恢复通信拓扑。
拓扑条件化判官 \TCJudge 随后结合轨迹、预测拓扑、
从独立标注语料中估计的经验失败先验，以及拓扑特有失败模式的简短描述，完成失败分类。
当编排结构固定时，恢复出的拓扑可以跨执行复用。}

Experimental results show a statistically significant association
between communication topology and failure type, with $\chi^2 = 409.9$
and $p = 1.2 \times 10^{-70}$. On the \num{851} MAST-clean traces,
ground-truth topology context raises gpt-mini's Macro-F1 from $0.173$
to $0.350$. With predicted topology, the pipeline achieves $0.346$,
approaching the trace-only gpt-5.4 baseline of $0.372$.
For \num{1000} traces under a fixed orchestration structure,
the projected pipeline cost, including one topology extraction, is
approximately $6\%$ of repeated gpt-5.4 diagnosis cost.
These results show that topology-conditioned context improves
failure diagnosis and supports lower-cost deployment.
\CNtranslation{摘要三}{%
实验结果表明，通信拓扑与失败类型之间存在统计显著关联，
其中 $\chi^2 = 409.9$，$p = 1.2 \times 10^{-70}$。
在包含 \num{851} 条轨迹的 MAST-clean 上，
提供真值拓扑上下文后，gpt-mini 的 Macro-F1 从 $0.173$ 提升至 $0.350$。
使用预测拓扑时，流水线取得 $0.346$，接近仅使用轨迹的 gpt-5.4 基线 $0.372$。
在固定编排结构下处理 \num{1000} 条轨迹时，
计入一次拓扑抽取的流水线成本估算约为重复使用 gpt-5.4 诊断的 $6\%$。
这些结果表明，拓扑条件化上下文能够改善失败诊断，并支持低成本部署。}
\end{abstract}


\maketitle
\thispagestyle{plain}

\section{Introduction}
\label{sec:intro}

\ifdefined\MAScopeBilingual
  \begin{figure}[!t]
\else
  \begin{figure}[!b]
\fi
  \centering
  \includegraphics[width=\linewidth]{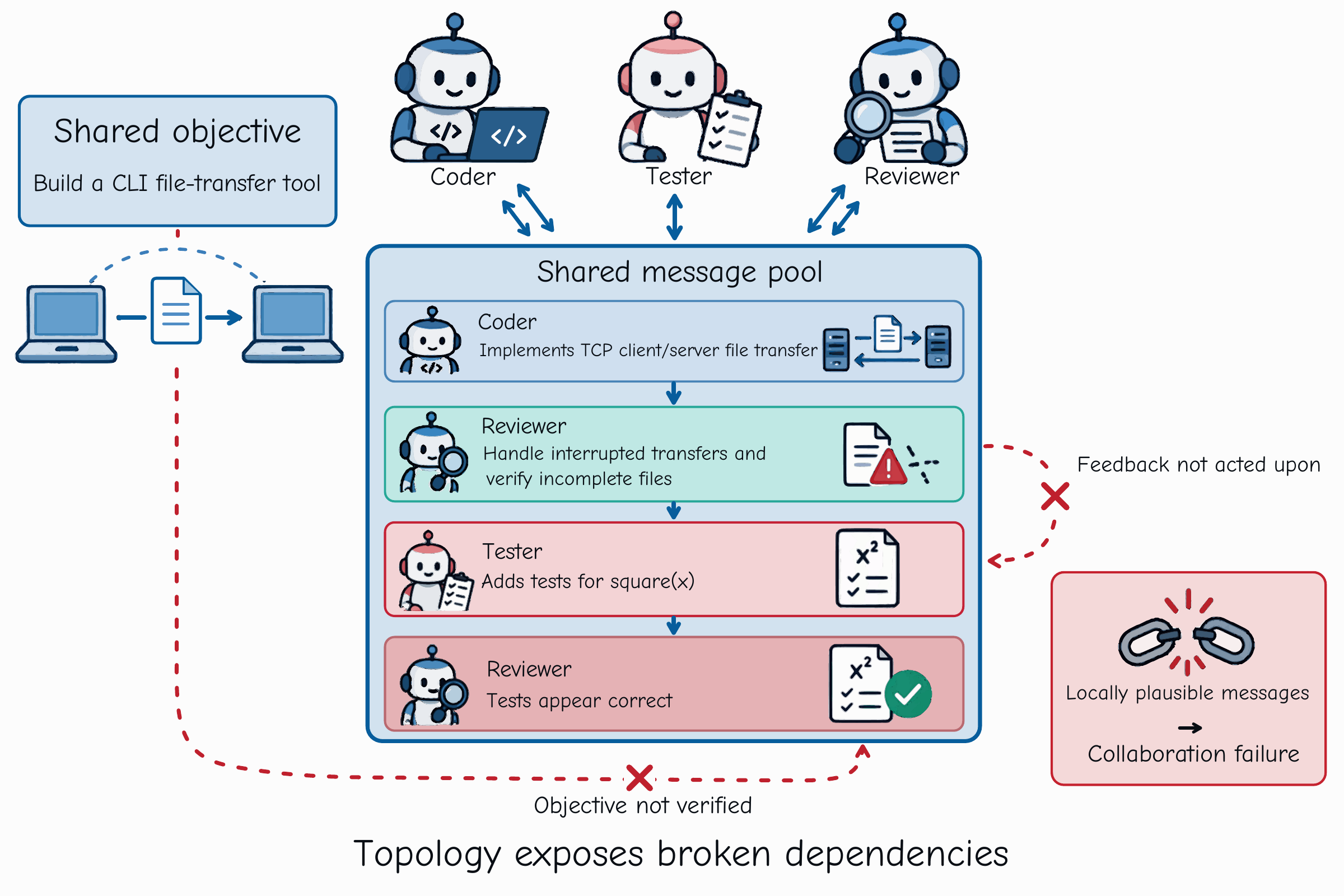}
  \caption{Shared message pool context guides checks of feedback use
  and task verification. In this trace, the reviewer's feedback was
  not acted upon and the requested verification was not performed.}
  \label{fig:motivating-case}
  \Description{A shared-message-pool execution in which a reviewer requests interrupted-transfer handling and verification, but the tester instead checks an unrelated square function and the next review accepts it. The topology view highlights the unaddressed feedback and missing task verification.}
\end{figure}

LLM multi-agent systems distribute complex tasks among agents that
plan, execute, and verify one another's work~\cite{li2023camel}. Frameworks such as
AutoGen, MetaGPT, and ChatDev support this division of
labour~\cite{wu2024autogen,hong2024metagpt,qian2024chatdev}.
Successful execution depends on how these agents coordinate as well
as on their individual capabilities. A requirement may be understood
by one agent yet disappear during a handoff, or several agents may
complete their subtasks while the final result remains unchecked.
Diagnosing such failures requires interpreting individual actions
within the relationships through which information is passed and
verified.

\CNtranslation{1 协作诊断动机}{%
LLM 多智能体系统将复杂任务分配给不同智能体，
由它们共同完成规划、执行与验证~\cite{li2023camel}。
AutoGen、MetaGPT 和 ChatDev 等框架支持这种任务分工
~\cite{wu2024autogen,hong2024metagpt,qian2024chatdev}。
任务能否成功，既取决于各智能体的能力，也取决于它们如何协作。
一项要求可能已被某个智能体理解，却在交接中丢失，
也可能多个智能体均完成了各自的子任务，最终结果却未经检查。
诊断这些失败，需要结合信息传递与验证关系来理解各智能体的行为。
}

Existing research analyses failures in multi-agent collaboration and
explores ways to improve collaborative performance.
MAST organises recurring problems into fourteen
failure modes~\cite{cemri2025mast}, while Who\&When and A2P examine
which agents and execution steps contribute to task
failure~\cite{zhang2025whowhen,west2025a2p}.
Other studies focus on the organisation of collaboration and use
graph-based methods to optimise communication topology for better
task performance and efficiency~\cite{zhuge2024gptswarm,zhang2024agentprune}.
These studies motivate examining the value of communication topology
as context for diagnosing collaboration failures.

\CNtranslation{1 研究问题}{%
已有研究分析多智能体协作中的失败，并探索改善协作表现的方法。
MAST 将反复出现的问题归纳为十四种失败模式~\cite{cemri2025mast}，
Who\&When 和 A2P 则考察哪些智能体及执行步骤导致任务失败
~\cite{zhang2025whowhen,west2025a2p}。
另一些研究关注协作的组织方式，
利用基于图的方法优化通信拓扑，以提高任务表现与效率
~\cite{zhuge2024gptswarm,zhang2024agentprune}。
这些研究促使我们考察通信拓扑作为协作失败诊断上下文的价值。
}

Communication topology describes information transfer relationships
among agents. Research on human organisations views coordination as
the management of dependencies among activities and links
organisational structure to the information processing required for
collaboration~\cite{galbraith1974organization,malone1994coordination},
while organisational structures in multi-agent systems likewise
influence agents' division of labour and information exchange
~\cite{horling2004organizational}. Recent work further shows
that agents that adhere rigidly to specialised roles may form a
collaboration topology composed of isolated groups, turning local
specialisation into bottlenecks at the system level
~\cite{panny2026specialists}.
Communication topology can provide structural context for diagnosing
collaboration failures, helping to relate individual actions to
information transfer among agents and shared task requirements.
\CNtranslation{1 组织结构视角}{%
通信拓扑描述智能体之间的信息传递关系。
人类组织研究将协调理解为对活动间依赖关系的管理
并将组织结构与协作所需的信息处理联系起来
~\cite{galbraith1974organization,malone1994coordination}，
多智能体系统中的组织结构同样影响智能体的分工与信息交换
~\cite{horling2004organizational}。
近期研究进一步表明，
严格固守专业化角色的智能体可能形成由孤立群体组成的协作拓扑，
使局部专业化转变为系统层面的瓶颈~\cite{panny2026specialists}。
通信拓扑能够为诊断协作失败提供结构上下文，
帮助将个体行为与智能体之间的信息传递及共享任务要求联系起来。}

Figure~\ref{fig:motivating-case} illustrates this perspective with a
representative execution from \textsf{MAST-clean}.
In a task to build a command-line file-transfer tool,
the reviewer requests support
for interrupted transfers and verification of incomplete files. The
tester instead adds tests for an unrelated square function, and the
next review accepts those tests as correct. In the shared message
pool, the reviewer's feedback is available to the tester and specifies
the task's verification requirements. \CompressionEdit{The final exchange is locally
coherent, but approval of an unrelated test neither addresses that
feedback nor verifies whether the requirements of the shared task
are met.}{The final exchange is locally coherent, but accepting an
unrelated test neither addresses the feedback nor verifies the shared
task's requirements.}
\CNtranslation{1 共享消息池案例}{%
图~\ref{fig:motivating-case} 通过 \textsf{MAST-clean}
中的一个代表性执行案例说明了这一视角。
在构建命令行文件传输工具的任务中，
审查员要求系统处理中断传输并验证不完整文件。
然而，测试员转而为无关的平方函数添加测试，
下一轮审查又认可了这些测试。
在共享消息池中，测试员能够获取审查员的反馈，
而且反馈明确了任务的验证要求。
\CompressionEdit{最后的交互在局部上连贯合理，
但认可无关测试既未落实这些反馈，也未验证共享任务的要求是否得到满足。
}{最后的交互在局部上连贯合理，但认可无关测试既未落实反馈，也未验证共享任务的要求。}
}

To incorporate communication structure into the diagnosis of failures
in multi-agent collaboration, we introduce \sysname{}, a two-stage
diagnostic framework. \CompressionEdit{Our empirical analysis finds that the distribution
of failure types differs across topologies, and \sysname{} accordingly
uses the failure patterns associated with each topology to guide
diagnosis.}{Our empirical analysis finds that failure-type distributions
differ across topologies, motivating topology-specific failure patterns
as diagnostic context.} In the first stage, the Trace Structural Extractor, \TSE{},
recovers communication relations and topology labels from message
evidence, providing structural context that heterogeneous logs may
not explicitly record. \CompressionEdit{In the second stage, the Topology-Conditioned
Judge, \TCJudge{}, is implemented with a lower-cost model and combines
the trace with the recovered topology and its associated failure
patterns to diagnose failures.}{In the second stage, the Topology-Conditioned
Judge, \TCJudge{}, uses a lower-cost model to diagnose failures from
the trace, recovered topology, and associated failure patterns.} \CompressionEdit{When the communication structure
remains stable, the recovered topology can be reused across executions,
reducing the cost of repeated diagnosis.}{When communication structure
remains stable, reusing topology across executions reduces repeated diagnosis cost.}

\CNtranslation{1 方法与复用}{%
为将通信结构纳入多智能体协作失败的诊断，
我们提出两阶段诊断框架 \sysname{}。
\CompressionEdit{实证分析发现，失败类型的分布因拓扑而异，
\sysname{} 据此利用各拓扑相关的失败模式来指导诊断。}{%
实证分析发现，失败类型的分布因拓扑而异，因此我们将拓扑特有的失败模式用作诊断上下文。}
在第一阶段，轨迹结构抽取器 \TSE{} 从消息证据中恢复通信关系和拓扑标签，
提供异构日志可能未显式记录的结构上下文。
\CompressionEdit{在第二阶段，拓扑条件化判官 \TCJudge{} 采用成本较低的模型，
结合执行轨迹、恢复出的拓扑及其相关失败模式进行失败诊断。}{%
在第二阶段，拓扑条件化判官 \TCJudge{} 使用成本较低的模型，
根据轨迹、恢复出的拓扑及相关失败模式诊断失败。}
\CompressionEdit{当通信结构保持稳定时，恢复出的拓扑可以跨执行复用，
从而降低重复诊断的成本。}{通信结构稳定时，跨执行复用拓扑可降低重复诊断成本。}
}

To evaluate \sysname{}'s diagnostic performance and deployment cost,
we assemble \textsf{MAScope-Bench} from natural execution traces,
traces with controlled fault injections, and traces from different
topologies within a single framework.
On \textsf{MAST-clean}, supplying ground-truth topology context raises
gpt-mini's Macro-F1 from $0.173$ to $0.350$.
Using topology recovered by \TSE{}, the complete gpt-mini pipeline
achieves a Macro-F1 of $0.346$, close to the $0.372$ achieved by
gpt-5.4 using only the trace.
\CompressionEdit{The controlled corpus shows that topology is still recoverable when
framework identity remains fixed, while injected traces show that
the diagnostic gains extend to additional framework and topology
settings.}{The controlled corpus shows topology recovery with framework
identity fixed, while injected traces extend diagnostic gains to
additional frameworks and topologies.} \CompressionEdit{For repeated diagnosis of \num{1000} traces under a fixed
topology, our cost projection includes one structural extraction and
estimates the gpt-mini pipeline's total cost at approximately $6\%$
of the cost of repeated gpt-5.4 diagnosis.}{For \num{1000} traces under
a fixed topology, the projected gpt-mini pipeline cost, including one
structural extraction, is approximately $6\%$ of repeated gpt-5.4 diagnosis cost.}
\CompressionEdit{These results \allowbreak{}show that explicit communication topology context
can narrow the diagnostic gap between models while reducing cost.}{These
results show that topology context can narrow diagnostic gaps between
models while reducing cost.}

\CNtranslation{1 结果概览}{%
为评估 \sysname{} 的诊断表现与部署成本，
我们整合自然执行轨迹、受控故障注入轨迹，
以及同一框架内不同拓扑的轨迹，构建 \textsf{MAScope-Bench}。
在 \textsf{MAST-clean} 上，提供真值拓扑上下文后，
gpt-mini 的 Macro-F1 从 $0.173$ 提升至 $0.350$。
使用 \TSE{} 恢复的拓扑时，完整的 gpt-mini 流水线取得 $0.346$ 的 Macro-F1，
接近 gpt-5.4 仅使用轨迹时的 $0.372$。
\CompressionEdit{控制语料表明，在框架身份保持固定时仍可恢复拓扑，
注入轨迹则表明诊断增益可以扩展至其他框架与拓扑设置。}{%
控制语料显示固定框架身份时可恢复拓扑，注入轨迹则将诊断增益扩展至其他框架与拓扑。}
\CompressionEdit{在固定拓扑下重复诊断 \num{1000} 条轨迹时，
我们的成本估算计入一次结构抽取，
估计 gpt-mini 流水线的总成本约为重复使用 gpt-5.4 诊断成本的 $6\%$。}{%
在固定拓扑下处理 \num{1000} 条轨迹时，计入一次结构抽取的
gpt-mini 流水线预估成本约为重复使用 gpt-5.4 诊断成本的 $6\%$。}
\CompressionEdit{这些结果表明，显式通信拓扑上下文可以缩小模型间的诊断差距，
并降低成本。}{这些结果表明，拓扑上下文可以缩小模型间的诊断差距并降低成本。}
}

The contributions of this work are as follows:
\begin{itemize}
  \item We develop \sysname{}, a two-stage framework that recovers
  communication topology from execution traces and uses the recovered
  topology and its associated failure patterns to guide a lower-cost
  model in diagnosing failures.
  \item We establish an empirical association between communication
  topology and failure patterns, identifying the failure types that
  are more prevalent under each topology.
  \item We construct \textsf{MAScope-Bench} with \num{2220} execution
  traces and supporting audit materials to evaluate topology recovery,
  failure diagnosis, transfer to unseen frameworks, and deployment cost.
\end{itemize}

\CNtranslation{1 贡献}{%
本文的贡献如下。
\begin{itemize}
  \item 我们提出两阶段框架 \sysname{}，从执行轨迹中恢复通信拓扑，
  并利用恢复出的拓扑及其相关失败模式，
  指导成本较低的模型进行失败诊断。
  \item 我们通过实证分析建立通信拓扑与失败模式之间的关联，
  识别各拓扑下更常见的失败类型。
  \item 我们构建包含 \num{2220} 条执行轨迹及配套审计材料的
  \textsf{MAScope-Bench}，用于评估拓扑恢复、失败诊断、
  向未见框架的迁移以及部署成本。
\end{itemize}
}


\section{Related Work}
\label{sec:related}

\subsection{Failure \allowbreak{}\allowbreak{}Diagnosis \allowbreak{}in LLM \allowbreak{}\allowbreak{}Multi-Agent Systems}
\label{sec:related:diagnosis}
\label{sec:related:attribution}
 Existing work on LLM multi-agent failure diagnosis mainly examines what types of failures occur and which agents or execution steps are responsible for them~\mbox{\cite{barke2026agentrx,li2026errorprobe}}.
\allowbreak{}Taxonomy-oriented work
 organizes recurring failure modes into structured categories.
MAST \mbox{\cite{cemri2025mast}}
 develops a taxonomy of 14 failure modes through expert analysis of execution traces and uses an LLM judge for scalable annotation.
 Other works categorize failures across agent modules~\mbox{\cite{zhu2025agentdebug}}
and explore
 formal dialogue verification~\mbox{\cite{nieves2026formaldialogue}}
and distributed constraint reasoning for LLM agents~\mbox{\cite{picard2026agenticdcr}}.
\allowbreak{}\allowbreak{}\allowbreak{}\allowbreak{}Localisation-oriented work
 focuses on identifying the agents and steps responsible for failures~\mbox{\cite{liu2026whowhenpro}}.
Who\&When \mbox{\cite{zhang2025whowhen}} formalises
 this task and provides annotated failure logs.
A2P \mbox{\cite{west2025a2p}}
 structures counterfactual reasoning for failure attribution, while
AgenTracer \mbox{\cite{zhang2025agentracer}}
trains an 8B tracer
 on trajectories annotated through counterfactual replay and fault injection.
 Other approaches use spectrum analysis~\mbox{\cite{ge2025famas}},
hierarchical context~\mbox{\cite{banerjee2025echo}},
or iterative judging~\mbox{\cite{zhu2026raffles}}
for failure attribution.
Related MARL work~\mbox{\cite{shefin2026marlforensics}}
analyses failure sources and propagation pathways.
 LumiMAS \mbox{\cite{lumimas2026}}
 combines system-level anomaly detection with vulnerability classification and root-cause analysis.
 These LLM-MAS diagnosis methods primarily rely on execution traces, without explicitly incorporating communication topology into the diagnostic input.
\CNtranslation{2.1 对照}{%
 已有的 LLM 多智能体失败诊断研究主要考察会发生哪些类型的失败，以及哪些智能体或执行步骤应对此负责~\mbox{\cite{barke2026agentrx,li2026errorprobe}}。
\allowbreak{}\emph{分类取向}\allowbreak{}的工作
 将反复出现的失败模式组织为结构化类别。
MAST~\mbox{\cite{cemri2025mast}}
 通过专家对执行轨迹的分析建立了包含十四种失败模式的分类学，并使用 LLM 判官支持大规模标注。
其他
研究按智能体模块对失败进行分类~\mbox{\cite{zhu2025agentdebug}}
，并探索面向 LLM 智能体的形式对话验证~\mbox{\cite{nieves2026formaldialogue}}
与分布式约束推理~\mbox{\cite{picard2026agenticdcr}}。
\allowbreak{}\allowbreak{}\emph{定位取向}\allowbreak{}的工作
 着重识别导致失败的智能体和步骤~\mbox{\cite{liu2026whowhenpro}}。
Who\&When~\mbox{\cite{zhang2025whowhen}}
 对这一任务进行形式化，并提供标注后的失败日志。
A2P~\mbox{\cite{west2025a2p}}
 通过结构化的反事实推理进行失败归因，而
AgenTracer~\mbox{\cite{zhang2025agentracer}}
 使用经反事实重放和故障注入标注的轨迹训练8B
追踪模型。
 其他方法采用谱分析~\mbox{\cite{ge2025famas}}、
层次化上下文~\mbox{\cite{banerjee2025echo}}
或迭代式判官~\mbox{\cite{zhu2026raffles}}进行失败归因。
相关 MARL 工作~\mbox{\cite{shefin2026marlforensics}}
分析失败源头及传播路径。
 LumiMAS~\mbox{\cite{lumimas2026}}
 结合系统层面的异常检测、漏洞分类和根因分析。
 这些 LLM 多智能体诊断方法主要依赖执行轨迹，未将通信拓扑显式纳入诊断输入。}

\subsection{Communication Topology of \allowbreak{}LLM \allowbreak{}MAS}
\label{sec:related:topology}
Communication topology describes how agents are connected and exchange information in LLM-MAS.
\allowbreak{}
Guo~et~al.\allowbreak{}~\mbox{\cite{guo2024llmmas}}
 distinguish four typical communication structures, namely layered, centralized, decentralized, and shared message pool.
 Existing research mainly constructs or adapts communication topologies to improve task performance and efficiency.
\allowbreak{}GPTSwarm \mbox{\cite{zhuge2024gptswarm}}
 optimises agent graphs at the node and edge levels, while
\allowbreak{}DyLAN \mbox{\cite{liu2024dylan}}
 adapts agent participation to each query.
\allowbreak{}AgentPrune \mbox{\cite{zhang2024agentprune}}
prunes redundant communication links to reduce token costs.
\allowbreak{}MaAS \mbox{\cite{zhang2025maas}} and AFlow \mbox{\cite{zhang2025aflow}}
 automate architecture and workflow search, respectively.
\allowbreak{}HyperAgent \mbox{\cite{zhang2026hyperagent}}
 optimises communication topology using hypergraphs.
 Generative approaches \mbox{\cite{jiang2025gdmtopology,leong2025amas}}
 produce task-specific topologies using
graph diffusion or trained graph designers.
\allowbreak{}G$^2$CP \mbox{\cite{benkhaled2026g2cp}}
 encodes messages as typed operations over a shared knowledge graph, while
\allowbreak{}ROUTER-HGC \mbox{\cite{xiao2026routerhgc}}
 selects collaboration modes and agent roles using a heterogeneous graph.
In MARL, ACORN \mbox{\cite{xie2025acorn}}
 reduces communication redundancy through acyclic coordination.
\allowbreak{}These studies show how communication design can improve task performance and efficiency~\mbox{\cite{zhu2025multiagentbench}}.
\CNtranslation{2.2 对照}{%
通信拓扑描述 LLM-MAS 中智能体的连接方式和信息交换方式。
\allowbreak{}
Guo~等人\allowbreak{}~\mbox{\cite{guo2024llmmas}}
 区分了四种典型通信结构，即分层式、中心式、去中心式和共享消息池。
 现有研究主要通过构建或调整通信拓扑，提高任务表现和效率。
\allowbreak{}GPTSwarm~\mbox{\cite{zhuge2024gptswarm}}
 在节点和边两个层面优化智能体图，而
\allowbreak{}DyLAN~\mbox{\cite{liu2024dylan}}
 根据每个查询动态调整智能体的参与。
\allowbreak{}AgentPrune~\mbox{\cite{zhang2024agentprune}}
 通过剪除冗余通信连接降低 token 成本。
\allowbreak{}MaAS~\mbox{\cite{zhang2025maas}} 与
AFlow~\mbox{\cite{zhang2025aflow}}
 分别自动进行智能体架构和工作流搜索。
\allowbreak{}HyperAgent~\mbox{\cite{zhang2026hyperagent}}
 使用超图优化通信拓扑。
 生成式方法通过图扩散或训练后的图设计器生成面向特定任务的拓扑~\mbox{\cite{jiang2025gdmtopology,leong2025amas}}。
\allowbreak{}G$^2$CP~\mbox{\cite{benkhaled2026g2cp}}
 将消息编码为共享知识图谱上的类型化操作，而
\allowbreak{}ROUTER-HGC~\mbox{\cite{xiao2026routerhgc}}
 使用异构图选择协作模式和智能体角色。
\allowbreak{}在 MARL 中， ACORN~\mbox{\cite{xie2025acorn}}
 通过无环协作减少通信冗余。
\allowbreak{}这些研究展示了通信设计如何改善任务表现和效率~\mbox{\cite{zhu2025multiagentbench}}。}

\subsection{\allowbreak{}From Execution Traces to Topology Labels}
\label{sec:related:recovery}
\label{sec:related:process}
\label{sec:related:framework}
 Recovering communication topology from execution traces remains underexplored in LLM-MAS.
 Classical process mining
 recovers workflow models from structured event logs~\mbox{\cite{vanderAalst2016process}}, whereas LLM-MAS traces
come from heterogeneous frameworks
 such as AutoGen, MetaGPT, ChatDev, LangGraph, CrewAI and Magentic.
Their native logging formats differ, and communication topology may not be explicitly recorded.
 AutoGen Studio~\mbox{\cite{dibia2024autogenstudio}} and LangSmith
 support trace inspection and debugging.
MAST \mbox{\cite{cemri2025mast}} and Who\&When \mbox{\cite{zhang2025whowhen}}
 provide failure annotations rather than communication topology labels.
 AgentPrune \mbox{\cite{zhang2024agentprune}}
 prunes communication graphs to reduce token costs rather than label coordination structures.
 \textbf{When the orchestration configuration is unavailable, topology must instead be inferred from observed agent interactions.}
 The remaining challenge is to recover reliable topology labels across frameworks from raw traces and ground those labels in message-level evidence.
\CNtranslation{2.3 对照}{%
 从执行轨迹恢复通信拓扑，在 LLM-MAS 中仍未得到充分研究。
 经典过程挖掘从结构化事件日志恢复工作流模型~\mbox{\cite{vanderAalst2016process}}，而
LLM-MAS 的轨迹则来自 AutoGen、\allowbreak{}MetaGPT、\allowbreak{}ChatDev、\allowbreak{}LangGraph、\allowbreak{}CrewAI 和 Magentic 等异构框架。
这些框架的原生日志格式各异，通信拓扑也未必得到显式记录。
 AutoGen Studio~\mbox{\cite{dibia2024autogenstudio}} 和 LangSmith 支持轨迹检查与调试。
MAST~\mbox{\cite{cemri2025mast}} 和 Who\&When~\mbox{\cite{zhang2025whowhen}}
 提供失败标注，而非通信拓扑标签。
AgentPrune~\mbox{\cite{zhang2024agentprune}}
 通过剪枝通信图降低 token 成本，而非进行协作结构标注。
 \textbf{在无法获得编排配置时，则需要从观测到的智能体交互中推断拓扑。}
 这里的挑战在于从不同框架的原始轨迹中恢复可靠的拓扑标签，并用消息层面的证据支撑这些标签。}

\section{Empirical Foundation: Topology
  \texorpdfstring{$\times$}{x} Failure Correlation}
\label{sec:empirical}

In this section, we examine the association between communication topology and failure patterns in LLM-MAS.
\CNtranslation{3 节总览}{%
本节中，\allowbreak{}我们考察通信拓扑与失败模式之间的关联，\allowbreak{}在 LLM-MAS 中。\allowbreak{}
}

\newcommand{\EmpiricalFloats}{%
\begin{figure}[!t]
  \centering
  \includegraphics[width=\linewidth]{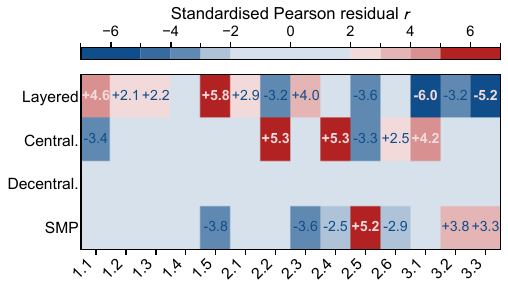}
  \ifdefined\MAScopeBilingual
    \caption{\DIFdel{Joint distribution of Guo four-class topology
    $\tau$ and the MAST-14 failure codes on MAST-clean (\num{851}
    traces, \num{3112} co-occurring $(\tau,\text{code})$ cells; each
    trace is a 14-way binary label). Cell colour is the standardised
    Pearson residual $(O-E)/\sqrt{E}$; cells with $|r|\!\geq\!2$ are
    annotated. Independence on the full $4\times 14$ table:
    $\chi^2 = 409.9$, $\mathrm{dof} = 39$, $p < 10^{-70}$. Each
    topology exhibits a distinct positive-residual fingerprint; the
    four anchor codes discussed in Section~\ref{sec:empirical:chi2}
    are the darkest red cells within each row.}
    \DIFadd{Topology--failure association on MAST-clean
    (\num{851} traces). Cells show standardised Pearson residuals for
    the $4\times 14$ contingency table, with $|r|\!\geq\!2$ annotated.
    The independence test gives $\chi^2=409.9$, $\mathrm{dof}=39$, and
    $p<10^{-70}$.}}
  \else
    \caption{Topology--failure association on MAST-clean
    (\num{851} traces). Cells show standardised Pearson residuals for
    the $4\times 14$ contingency table, with $|r|\!\geq\!2$ annotated.
    The independence test gives $\chi^2=409.9$, $\mathrm{dof}=39$, and
    $p<10^{-70}$.}
  \fi
  \label{fig:chi2-heatmap}
  \Description{A four-by-fourteen heat map of standardized Pearson residuals between communication topology and MAST failure codes. Each topology row has a different cluster of positive residuals, with Layered associated with information loss, Centralized with verification misalignment, Decentralized with coordination stall, and Shared Message Pool with semantic drift.}
\end{figure}

\begin{figure*}[!t]
  \centering
  \includegraphics[width=0.95\textwidth]{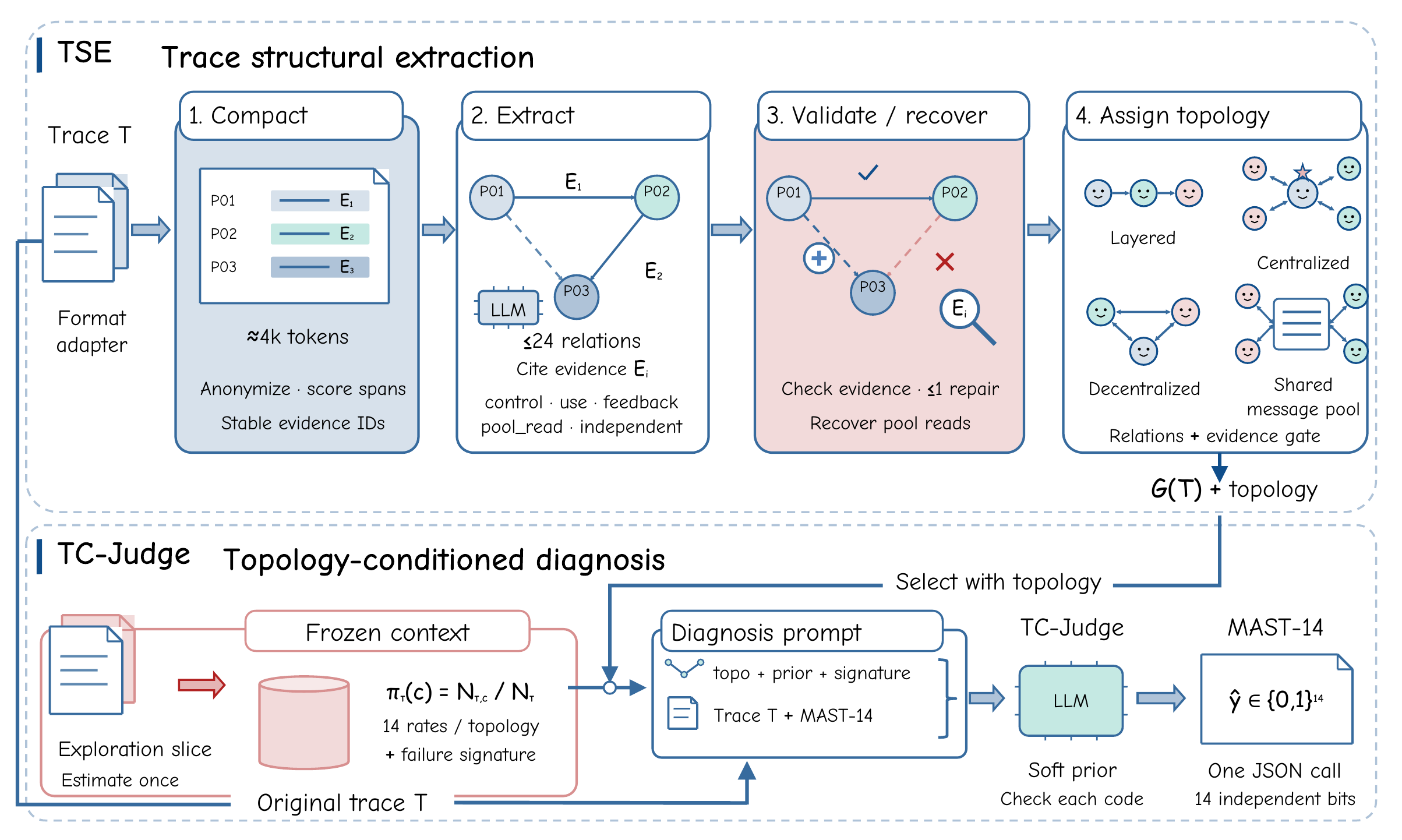}
  \ifdefined\MAScopeBilingual
    \caption{\DIFdel{\sysname{} two-stage pipeline. \TSE transforms a
    heterogeneous framework trace into a canonical relation graph $G(T)$
    and topology label $\hat\tau$ through compaction, evidence-grounded
    relation extraction, validation and recovery, and structural
    labelling. The recovered topology selects a frozen empirical prior
    $\pi_{\hat\tau}$ and failure signature estimated once from a disjoint
    exploration slice. \TCJudge combines this context with the original
    trace and emits a fourteen-dimensional binary diagnosis
    $\hat{\mathbf{y}} \in \{0,1\}^{14}$ over the MAST taxonomy.}
    \DIFadd{\sysname{} pipeline. \TSE converts a heterogeneous trace into
    an evidence-grounded relation graph and topology label $\hat\tau$.
    \TCJudge combines the trace, predicted topology, frozen topology prior,
    and failure signature to output a MAST-14 diagnosis.}}
  \else
    \caption{\sysname{} pipeline. \TSE converts a heterogeneous trace into
    an evidence-grounded relation graph and topology label $\hat\tau$.
    \TCJudge combines the trace, predicted topology, frozen topology prior,
    and failure signature to output a MAST-14 diagnosis.}
  \fi
  \label{fig:pipeline}
  \Description{A two-stage pipeline. The first stage adapts and compacts a heterogeneous trace, extracts and validates evidence-grounded relations, and assigns a topology label. The second stage combines the original trace with the topology label, a topology-specific prior, and a failure signature to produce a fourteen-bit MAST diagnosis.}
\end{figure*}
}
\ifdefined\MAScopeBilingual
  \EmpiricalFloats
\fi

\subsection{Setting}
\label{sec:empirical:setting}
\ifdefined\MAScopeBilingual\else
\begin{table}[!b]
  \centering
  \small
  \setlength{\tabcolsep}{4pt}
  \caption{ \textsf{MAScope-Bench} composition.  Lay., Cen., Dec., and SMP denote \Layered, \Centralized, \Decentralized, and \SharedPool, based on each declared orchestration graph.    }
  \label{tab:fingerprint}
  \begin{tabular}{@{}lccccc@{}}
    \toprule
    Corpus & \# traces & Lay. & Cen. & Dec. & SMP \\
    \midrule
    \textsf{MAST-clean}   & \num{851} & \num{256} & \num{165} & \num{0}   & \num{430} \\
    \textsf{LG-Ctrl}      & \num{408} & \num{102} & \num{102} & \num{102} & \num{102} \\
    \textsf{Inject-Mix}   & \num{961} & \num{170} & \num{284} & \num{507} & \num{0}   \\
    \bottomrule
  \end{tabular}

\end{table}

\fi
A raw multi-agent execution trace is a tuple $T = (A, M, R, S)$\allowbreak{}\allowbreak{}, where \allowbreak{}$A$ is
 the set of agents with declared roles, $M$ is the time-stamped message
 stream $(t_i, \allowbreak{}\mathrm{sender}_i, \allowbreak{}\mathrm{receiver}_i,
\allowbreak{}\mathrm{content}_i)$, $R$ collects tool and retrieval returns, and $S$
carries external state\allowbreak{}\allowbreak{}. \allowbreak{}From $T$ we derive a canonical \allowbreak{}\allowbreak{}DAG
\allowbreak{} $G(T) = (V, E)$, \allowbreak{}\allowbreak{}whose \allowbreak{}vertices \allowbreak{}\allowbreak{}are agent turns and whose \allowbreak{}\allowbreak{}edges \allowbreak{}\allowbreak{}are
 evidence-carrying parent--child dependencies\allowbreak{}\allowbreak{}. The associated topology label
$\tau \in \{\Layered,\allowbreak{} \Centralized,\allowbreak{} \Decentralized,\allowbreak{} \SharedPool\}$
 follows Guo et al.'s four-class taxonomy\allowbreak{}~\mbox{\cite{guo2024llmmas}}\allowbreak{}.
 Failure annotations are represented by $\mathbf{y} \in \{0,1\}^{14}$ using the MAST taxonomy~\mbox{\cite{cemri2025mast}}, which covers \emph{Agent-Level Specification}, \emph{Inter-Agent Misalignment}, and \emph{Task Verification}. A trace may have multiple failure codes. For descriptive analysis, we group withheld or ignored information as \InfoFailure{}, role violations as \CollabStall{}, and specification violations, inconsistency, or reasoning--action mismatch as \SemanticDrift{}. Task-verification failures form the \VerifyMisalign{} group.
\CNtranslation{3.1 设定}{%
一条原始多智能体执行轨迹是元组 $T = (A, M, R, S)$，\allowbreak{}其中 $A$ 是具有已声明角色的智能体集合，\allowbreak{}$M$ 是带时间戳的消息流 $(t_i, \allowbreak{}\mathrm{sender}_i, \allowbreak{}\mathrm{receiver}_i, \allowbreak{}\mathrm{content}_i)$，\allowbreak{}$R$ 汇总工具与检索返回，\allowbreak{}$S$ 承载外部状态。\allowbreak{}
从 $T$ 中，\allowbreak{}我们导出规范有向无环图 $G(T) = (V, E)$，\allowbreak{}其顶点是智能体轮次，\allowbreak{}并且其边是携带证据的父子依赖。\allowbreak{}相关的拓扑标签 $\tau \in \{\Layered,\allowbreak{} \Centralized,\allowbreak{} \Decentralized,\allowbreak{} \SharedPool\}$ 遵循 Guo 等人的四类分类体系~\mbox{\cite{guo2024llmmas}}。\allowbreak{}
失败标注被表示为 $\mathbf{y} \in \{0,1\}^{14}$，\allowbreak{}采用 MAST 分类体系~\mbox{\cite{cemri2025mast}}，\allowbreak{}该体系涵盖\emph{智能体级规范}、\allowbreak{}\emph{智能体间失配}和\emph{任务验证}。\allowbreak{}
一条轨迹可以具有多个失败码。\allowbreak{}
为进行描述性分析，\allowbreak{}我们将被隐瞒或被忽略的信息归为信息丢失，\allowbreak{}将角色违背归为协作停滞，\allowbreak{}将规范违背、\allowbreak{}不一致或推理与行动不一致归为语义漂移。\allowbreak{}任务验证失败构成验证失配组。\allowbreak{}
}

\subsection{Corpora and Ground Truth}
\label{sec:empirical:corpora}
We use three corpora summarised in Table~\ref{tab:fingerprint}, with construction and auditing detailed in Section~\ref{sec:experiments:dataset}.
\emph{MAST-clean} contains \num{851} traces from the MAST release~\cite{cemri2025mast}, retaining the original MAST-14 failure annotations, and we independently relabel a stratified subset of 200 traces.
\emph{Inject-Mix} contains \num{961} traces constructed using controlled fault injection across Layered, Centralized, and Decentralized orchestration variants, covering the Decentralized topology absent from MAST-clean and providing external validation across runtimes.
The \num{408} traces in \emph{LG-Ctrl} are controlled end to end by a single LangGraph harness in which all four topologies share the same \texttt{StateGraph} codebase and topology-agnostic agent identifiers, so any observed topology--failure association cannot be attributed to framework fingerprints.
Across all three corpora the topology label $\tau$ is obtained from the original framework-declared orchestration graph and used as ground truth for evaluating topology extraction.
\CNtranslation{3.2 语料与真值}{%
我们使用三套语料，\allowbreak{}汇总于表~\ref{tab:fingerprint}，\allowbreak{}构建与审计详见第~\ref{sec:experiments:dataset}~节。\allowbreak{}
\emph{MAST-clean} 包含 \num{851} 条轨迹，\allowbreak{}来自 MAST 发布集~\cite{cemri2025mast}，\allowbreak{}保留原始 MAST-14 失败标注，\allowbreak{}而且我们独立重标一个包含\allowbreak{}200\allowbreak{}条轨迹的分层子集。\allowbreak{}
\emph{Inject-Mix} 包含 \num{961} 条轨迹，\allowbreak{}这些轨迹使用受控故障注入构建，\allowbreak{}跨越分层、\allowbreak{}中心化和去中心化编排变体，\allowbreak{}覆盖 MAST-clean 中缺失的去中心化拓扑，\allowbreak{}并提供跨运行环境的外部验证。\allowbreak{}
\emph{LG-Ctrl} 中的 \num{408} 条轨迹由单一 LangGraph 测试框架端到端控制，\allowbreak{}其中四种拓扑共享同一份 \texttt{StateGraph} 代码库和与拓扑无关的智能体标识符，\allowbreak{}因此任何观测到的拓扑与失败关联都不能归因于框架指纹。\allowbreak{}
在全部三套语料中，\allowbreak{}拓扑标签 $\tau$ 取自原始框架声明的编排图，\allowbreak{}并被用作评估拓扑提取的真值。\allowbreak{}
}

\ifdefined\MAScopeBilingual\else
  \EmpiricalFloats
\fi

\subsection{Independence Test and Per-Topology Anchors}
\label{sec:empirical:chi2}
\label{sec:empirical:fingerprint}
We construct a contingency table of topology and MAST failure annotations on MAST-clean to test their independence. As shown in Figure~\ref{fig:chi2-heatmap}, the test yields $\chi^2 = 409.9$ and $p = 1.2 \times 10^{-70}$, indicating a significant association between topology and failure distributions. The standardised Pearson residuals further show that positive residuals concentrate on different failure types across topologies, revealing characteristic failure patterns for each topology.
\CNtranslation{3.3 独立性检验}{%
我们构建拓扑与 MAST 失败标注的联合表，在 MAST-clean 上检验两者的独立性。
如图~\ref{fig:chi2-heatmap} 所示，检验得到 $\chi^2 = 409.9$ 和 $p = 1.2 \times 10^{-70}$，
表明拓扑与失败分布之间存在显著关联。
标准化 Pearson 残差进一步显示，不同拓扑的正残差集中于不同失败类型，
从而揭示每种拓扑特有的失败模式。}

\CompressionEdit{%
Specifically, positive residuals for \Layered{}, \Centralized{}, \Decentralized{}, and \SharedPool{} topologies concentrate on \InfoFailure{}, \VerifyMisalign{}, \CollabStall{}, and \SemanticDrift{}, respectively. In \Layered{} pipelines, \InfoFailure{} involves $2.4$ and $2.5$ for information withholding and ignored input, which manifest as the loss of upstream facts before they reach downstream stages along the forward-only communication path. In \Centralized{} topologies, \VerifyMisalign{} involves $3.1$ through $3.3$, with a hub performing global verification whose criterion gradually becomes misaligned with the workers it supervises. In \Decentralized{} topologies, \CollabStall{} manifests as role violations under $2.1$, and symmetric peers bearing full responsibility for coordination enter cycles of mutual waiting or role confusion. In \SharedPool{} topologies, \SemanticDrift{} involves $2.2$, $2.3$, and $2.6$, corresponding to specification violations, inconsistency, and mismatches between reasoning and action, respectively. The blackboard accumulates context with ambiguous scope, and later readers silently reinterpret earlier intent. Together, these patterns connect coordination structures to concrete failures, providing a failure anchor for each topology and a common basis for comparing failure distributions across corpora.
}{%
In \Layered{} pipelines, \InfoFailure{} involves $2.4$ and $2.5$ for information withholding and ignored input, as upstream facts are lost before reaching downstream stages along the forward-only communication path. In \Centralized{} topologies, \VerifyMisalign{} involves $3.1$ through $3.3$, as the hub's global verification criterion gradually becomes misaligned with its workers. In the supplementary corpora's \Decentralized{} traces, \CollabStall{} manifests as role violations under $2.1$, with symmetric peers responsible for coordination entering cycles of mutual waiting or role confusion. In \SharedPool{} topologies, \SemanticDrift{} involves $2.2$, $2.3$, and $2.6$ for specification violations, inconsistency, and reasoning--action mismatch. The blackboard accumulates context with ambiguous scope, and later readers silently reinterpret earlier intent. These structural failure anchors provide a common basis for cross-corpus comparison.}
\CNtranslation{3.3 逐拓扑失败锚点}{%
\CompressionEdit{%
具体而言，分层、中心化、去中心化和共享消息池拓扑的正残差分别集中于信息丢失、验证失配、协作停滞和语义漂移。
在分层流水线中，信息丢失涉及 $2.4$ 和 $2.5$，对应信息隐瞒和被忽略的输入，
它们表现为上游事实沿仅向前的通信路径到达下游阶段之前丢失。
在中心化拓扑中，验证失配涉及 $3.1$ 至 $3.3$，
中心节点执行全局验证，而验证标准逐渐与该节点监督的工作智能体失配。
在去中心化拓扑中，协作停滞表现为 $2.1$ 下的角色违背，
承担全部协调责任的对称对等智能体进入相互等待或角色混淆的循环。
在共享消息池拓扑中，语义漂移涉及 $2.2$、$2.3$ 和 $2.6$，
分别对应规范违背、不一致以及推理与行动不一致。
黑板积累作用域模糊的上下文，后续读取者悄然重新解释先前意图。
共同来看，这些模式将协作结构与具体失败联系起来，
为每种拓扑提供失败锚点，并为跨语料比较失败分布提供共同依据。}{%
在分层流水线中，信息丢失涉及 $2.4$ 和 $2.5$，对应信息隐瞒和被忽略的输入，表现为上游事实沿仅向前的通信路径到达下游阶段之前丢失。
在中心化拓扑中，验证失配涉及 $3.1$ 至 $3.3$，表现为中心节点的全局验证标准逐渐与其监督的工作智能体失配。
在补充语料的去中心化轨迹中，协作停滞表现为 $2.1$ 下的角色违背，负责协调的对称对等智能体进入相互等待或角色混淆的循环。
在共享消息池拓扑中，语义漂移涉及 $2.2$、$2.3$ 和 $2.6$，对应规范违背、不一致以及推理与行动不一致。
黑板积累作用域模糊的上下文，后续读取者悄然重新解释先前意图。
这些结构性失败锚点为跨语料比较提供共同依据。}}

The associations between the four topologies and their characteristic failure patterns remain qualitatively consistent across MAST-clean, LG-Ctrl, and Inject-Mix. The leave-one-framework-out evaluation further shows that the diagnostic method retains $82\%$ of its in-distribution Macro-F1. Along with the independence test and residual analysis, these results inform the inclusion of topology in \TCJudge{}'s zero-shot prompt.
\CNtranslation{3.3 跨语料一致性}{%
这四类拓扑与其特有失败模式之间的关联在 MAST-clean、LG-Ctrl 和 Inject-Mix 中保持定性一致。
留一框架评测进一步显示，诊断方法保留其同分布 Macro-F1 的 $82\%$。
连同独立性检验和残差分析，这些结果为将拓扑纳入 \TCJudge{} 的零样本提示提供参考。}


\section{Two-Stage Diagnosis Pipeline}
\label{sec:pipeline}

\sysname{} follows the two-stage pipeline shown in Figure~\ref{fig:pipeline}, first recovering the collaboration topology and then using topology information to diagnose failures in execution traces. We make this separation because topology $\tau$ and failure labels $\mathbf{y}$ change at different rates. For example, a deployed layered LangGraph pipeline retains the same topology across thousands of runs, while failure labels vary with each execution. These two stages are implemented by the Trace Structural Extractor \TSE{} and the Topology-Conditioned Judge \TCJudge{}, respectively. \TSE{} recovers a canonical relation graph $G(T)$ and the Guo topology label $\hat\tau$ from a heterogeneous framework trace $T$, and \TCJudge{} combines the original trace $T$ with this label $\hat\tau$ to produce the MAST-14 binary failure vector $\hat{\mathbf{y}} \in \{0,1\}^{14}$. Reusing the recovered topology within a fixed deployment amortises extraction across its executions, leaving \TCJudge{} as the marginal runtime cost for each trace.
\CNtranslation{4 章总览}{%
\sysname{} 采用图~\ref{fig:pipeline} 所示的两阶段流水线。系统首先恢复协作拓扑，随后利用拓扑信息诊断执行轨迹中的失败。我们进行这一拆分，是因为拓扑 $\tau$ 与失败标签 $\mathbf{y}$ 的变化速率不同。例如，一个已部署的分层 LangGraph 流水线可以在数千次运行中保持相同拓扑，而失败标签会随每次执行而变化。这两个阶段分别由轨迹结构提取器 \TSE{} 和拓扑条件化判官 \TCJudge{} 实现。\TSE{} 从异构框架轨迹 $T$ 中恢复规范关系图 $G(T)$ 与 Guo 拓扑标签 $\hat\tau$。\TCJudge{} 将原始轨迹 $T$ 与标签 $\hat\tau$ 结合，生成 MAST-14 二元失败向量 $\hat{\mathbf{y}} \in \{0,1\}^{14}$。在固定部署中复用已恢复的拓扑，可以将提取成本分摊到多次执行中，使 \TCJudge{} 成为处理每条轨迹时的边际运行成本。}

\subsection{Stage 1. \texorpdfstring{\TSE}{TSE}, Trace Structural Extraction}
\label{sec:pipeline:tse}
\TSE recovers the relation graph $G(T)$ and topology label $\hat\tau$ from trace $T$ by grounding relations and topology labels in identifiable trace evidence. Its four processing stages cover compaction, relation extraction, validation with repair, and structural labelling. Deterministic rules support speaker attribution over long distances and recovery of implicit shared-channel reads, giving the LLM explicit evidence for relations that require context across a long trace. All stages use a common schema consisting of a speaker-tagged message stream and optional shared-channel markers. A thin format adapter connects each framework to this schema, and we provide adapters for LangGraph, AutoGen, CrewAI, ChatDev, MetaGPT, Magentic, and Simple Chain. This shared representation lets the extraction process begin with the same evidence preparation across all seven frameworks.
\CNtranslation{4.1 \TSE 概览}{%
\TSE 通过将关系与拓扑标签建立在可识别的轨迹证据上，从轨迹 $T$ 中恢复关系图 $G(T)$ 和拓扑标签 $\hat\tau$。它的四个处理阶段涵盖压缩、关系提取、带修复的校验以及结构标注。确定性规则支持跨长距离的说话者归因和隐式共享通道读取恢复，从而为需要长轨迹上下文的关系向 LLM 提供显式证据。所有阶段使用同一种数据模式，其中包含带说话者标签的消息流和可选的共享通道标记。轻量格式适配器将每个框架连接到该数据模式。我们为 LangGraph、AutoGen、CrewAI、ChatDev、MetaGPT、Magentic 和 Simple Chain 提供适配器。这种共享表示使七个框架的提取过程都能从相同的证据准备步骤开始。}

\ifdefined\MAScopeBilingual\else
\begin{table*}[!t]
  \centering
  \begin{minipage}{0.90\textwidth}
    \centering
    \small
    \setlength{\tabcolsep}{6pt}
    \renewcommand{\arraystretch}{1.04}
    \caption{ Zero-shot diagnostic results on \textsf{MAST-clean}.  No method uses its failure labels for fitting.  The final column reports the gain from $M_0$ to $M_4$ on the same  model. }
    \label{tab:m0m4-indist}
    \begin{tabular*}{\linewidth}{@{\extracolsep{\fill}}lllrr@{}}
      \toprule
      Family & Method & Setting & Macro-F1 & Gain over matched $M_0$ \\
      \midrule
      External protocol & MAST-Judge & gpt-mini & 0.131 & -- \\
      External protocol & MAST-Judge & gpt-5.4 & 0.363 & -- \\
      \addlinespace[2pt]
      Matched control & $M_0$ & gpt-mini & 0.173 & Reference \\
      Matched control & $M_0$ & gpt-5.4 & 0.372 & Reference \\
      \addlinespace[2pt]
      \textbf{Ours} & $M_4$ & gpt-mini with true topology
        & \textbf{0.350} & \textbf{$+0.177$ and $+102\%$} \\
      \textbf{Ours} & $M_4$ & gpt-5.4 with true topology
        & \textbf{0.399} & \textbf{$+0.027$ and $+7.3\%$} \\
      \bottomrule
    \end{tabular*}
  \end{minipage}
\end{table*}

\fi

\TSE runs four stages per trace.
\textbf{(i) Compact and anonymize.} Compaction prepares a bounded evidence view while preserving the participants and local cues needed for relation extraction. We parse $T$ into a speaker-tagged message stream and replace participant names with placeholders $P_{01}, P_{02}, \ldots$ in order of first appearance. Sentence-level spans are scored using named co-mentions, interaction verbs, shared-channel markers, and words indicating independent initiation. \CompressionEdit{The selected spans form a view of approximately 4k tokens, and each retains a stable identifier $E_i$. These identifiers allow the extractor to cite the evidence for each relation in the next stage.}{The selected spans form a view of approximately 4k tokens, each retaining a stable identifier $E_i$ for citing relation evidence.}
\CNtranslation{4.1 \TSE 第一步}{%
\TSE 对每条轨迹执行四个阶段。\textbf{第一步是压缩与匿名化。} 压缩在保留关系提取所需参与者与局部线索的同时，构造一个长度受限的证据视图。我们将 $T$ 解析为带说话者标签的消息流，并按照首次出现顺序将参与者名称替换为占位符 $P_{01}, P_{02}, \ldots$。系统根据名称共同出现、交互动词、共享通道标记以及表示独立发起的词语，对句子级片段进行评分。\CompressionEdit{入选片段构成约四千 token 的视图，并各自保留稳定标识符 $E_i$。这些标识符使提取器能够在下一阶段引用每项关系的证据。}{入选片段构成约四千 token 的视图，每个片段保留稳定标识符 $E_i$，用于引用关系证据。}}

\textbf{(ii) Extract.} Relation extraction turns the retained evidence into a compact set of interaction candidates. The LLM returns at most 24 relations, each assigned one of the five types $\mathtt{control}$, $\mathtt{use}$, $\mathtt{feedback}$, $\mathtt{pool\_read}$, and $\mathtt{independent}$. Each candidate cites one or two spans $E_i$ that identify the speaker and, where applicable, the counterpart and store. \CompressionEdit{The resulting relation and evidence pairs provide the validator with a concrete basis for checking each candidate.}{}
\CNtranslation{4.1 \TSE 第二步}{%
\textbf{第二步是提取。} 关系提取将保留的证据转换为紧凑的交互候选集合。LLM 最多返回二十四项关系，每项关系被赋予五种类型之一，即 $\mathtt{control}$、$\mathtt{use}$、$\mathtt{feedback}$、$\mathtt{pool\_read}$ 或 $\mathtt{independent}$。每个候选关系引用一到两个片段 $E_i$，用于标识说话者，并在适用时标识交互对象与存储位置。\CompressionEdit{由此得到的关系与证据对为校验器检查每个候选关系提供具体依据。}{}}

\textbf{(iii) Validate and recover relations.} Validation and recovery assemble the relation set used for structural labelling. A rule-based validator checks each candidate against its cited span for JSON schema conformance, direction, grounding within the same message, a named counterpart, and a named store. If the rejection rate exceeds a threshold or shared-channel markers accompany a sparse graph, one bounded LLM repair call retries the rejected candidates through the same validator. A regular expression with a narrow context then supplements free-form generation by recovering implicit pool reads. \CompressionEdit{The validated and recovered relations together supply the evidence for assigning a topology.}{}
\CNtranslation{4.1 \TSE 第三步}{%
\textbf{第三步是校验与关系恢复。} 校验和恢复过程组装用于结构标注的关系集合。基于规则的校验器根据候选关系引用的片段，检查 JSON 数据模式一致性、方向、同一消息内的证据落点、具名交互对象以及具名存储位置。如果拒绝率超过阈值，或共享通道标记与稀疏图同时出现，系统会进行一次受限的 LLM 修复调用，并让修复后的候选关系再次通过同一校验器。随后，一个采用窄上下文的正则表达式通过恢复隐式消息池读取来补充自由生成结果。\CompressionEdit{通过校验和恢复得到的关系共同构成拓扑分配的证据。}{}}

\textbf{(iv) Assign the topology label.} Structural labelling converts these relations into a Guo topology label through a classifier and a positive-evidence gate. Relations are aggregated from message speakers $P_i$ to graph nodes $N_j$ and serialised into a table consisting entirely of these aggregated relations. This table is the complete input to the structural-labelling LLM, which assigns one of the four Guo classes or abstains. \CompressionEdit{Its role is to classify collaboration structure, while \TCJudge subsequently classifies failures.}{} A deterministic gate resolves the final label by giving highest priority to $\mathtt{pool\_read}$ relations with a named producer. Independent or mutual activity comes next, followed by a persistent delegation chain and then unidirectional information flow. \TSE returns the canonical DAG $G(T)$ and $\hat\tau$ together, with every relation and the final label linked to named spans in the retained evidence.
\CNtranslation{4.1 \TSE 第四步}{%
\textbf{第四步是分配拓扑标签。} 结构标注通过分类器和正证据门控，将这些关系转换为 Guo 拓扑标签。系统将消息说话者 $P_i$ 的关系聚合到图节点 $N_j$，并将其序列化为一个完全由聚合关系组成的表格。该表格是结构标注 LLM 的完整输入。模型分配四种 Guo 类别之一，或选择弃权。\CompressionEdit{它负责分类协作结构，而 \TCJudge 随后负责分类失败。}{}确定性门控通过证据优先级确定最终标签。带有具名生产者的 $\mathtt{pool\_read}$ 关系优先级最高，其次是独立或互惠活动，再其次是持续委派链，最后是单向信息流。\TSE 同时返回规范 DAG $G(T)$ 和 $\hat\tau$，其中每项关系与最终标签都链接到保留证据中的具名片段。}

\subsection{Stage 2. \texorpdfstring{\TCJudge}{TC-Judge}, Topology-Conditioned LLM-Judge}
\label{sec:pipeline:tcjudge}
\TCJudge{} uses the recovered topology to interpret failure evidence in each execution trace. Given $T$ and $\hat\tau$, it produces $\hat{\mathbf{y}} \in \{0,1\}^{14}$ in one structured JSON call, with one independently assigned bit for each MAST failure code. Topology enters its prompt through the label $\hat\tau$, a per-topology empirical prior $\pi_{\hat\tau}$, and a short failure signature describing characteristic interactions. With this topology context, the judge performs zero-shot diagnosis with fixed model parameters.
\CNtranslation{4.2 \TCJudge 概览}{%
\TCJudge{} 使用恢复的拓扑解释每条执行轨迹中的失败证据。给定 $T$ 和 $\hat\tau$，它通过一次结构化 JSON 调用生成 $\hat{\mathbf{y}} \in \{0,1\}^{14}$，并为每个 MAST 失败码独立分配一个比特。拓扑通过标签 $\hat\tau$、逐拓扑经验先验 $\pi_{\hat\tau}$ 以及描述特征性交互的简短失败签名进入提示。借助这一拓扑背景，判官在模型参数固定的条件下执行零样本诊断。}

\TCJudge{} operates in two steps.
\textbf{(i) Topology context preparation.} For each topology $\tau$, we prepare an empirical prior and a short failure-signature paragraph. For each of the fourteen MAST codes $c$, the occurrence rate is computed as $\pi_\tau(c)=\frac{N_{\tau,c}}{N_\tau}$, where $N_\tau$ counts traces with topology $\tau$ and $N_{\tau,c}$ counts those also labelled with code $c$. Using a 400-trace exploration slice drawn from an earlier data release, we estimate these rates once and then freeze them. The 851-trace MAST-clean evaluation set comes from a later release and is disjoint from the exploration slice. The failure-signature paragraph describes characteristic roles and interaction patterns under $\tau$. Together with the topology label, the prior and signature form the context used in the diagnostic prompt.
\CNtranslation{4.2 \TCJudge 第一步}{%
\TCJudge{} 分两步运行。\textbf{第一步是拓扑背景准备。} 对每种拓扑 $\tau$，我们准备一个经验先验和一段简短的失败签名。对于十四个 MAST 码中的每个码 $c$，其出现率计算为 $\pi_\tau(c)=\frac{N_{\tau,c}}{N_\tau}$。其中 $N_\tau$ 表示拓扑为 $\tau$ 的轨迹数量，$N_{\tau,c}$ 表示其中同时标有代码 $c$ 的轨迹数量。我们使用从较早数据版本抽取的四百条探索切片一次性估计这些比率，随后将其冻结。包含八百五十一条轨迹的 MAST-clean 评测集来自较晚版本，与探索切片互不重叠。失败签名段落描述拓扑 $\tau$ 下的特征性角色和交互模式。拓扑标签、先验与签名共同构成诊断提示所使用的背景。}

\textbf{(ii) Topology-conditioned diagnosis.} The diagnostic prompt combines the trace $T$ with the recovered topology label $\hat\tau$, the corresponding prior $\pi_{\hat\tau}$, and the failure-signature paragraph. The prior component consists entirely of the aggregate rate for each code under $\hat\tau$, with all fourteen rates presented together. A soft-prior calibration instruction directs the judge to check each code independently against the trace evidence and then assign its corresponding bit. The judge returns the resulting fourteen-bit vector $\hat{\mathbf{y}}$ in one structured JSON call.
\CNtranslation{4.2 \TCJudge 第二步}{%
\textbf{第二步是拓扑条件化诊断。} 诊断提示将轨迹 $T$ 与恢复的拓扑标签 $\hat\tau$、对应先验 $\pi_{\hat\tau}$ 以及失败签名段落相结合。先验部分仅包含 $\hat\tau$ 下每个失败码的聚合比率，并同时呈现全部十四项比率。软先验校准指令要求判官针对轨迹证据独立检查每个失败码，随后分配对应比特。判官通过一次结构化 JSON 调用返回所得的十四比特向量 $\hat{\mathbf{y}}$。}

\textbf{Remark 1. Zero-shot design.} The zero-shot design supports diagnosis under the uneven label coverage of MAST-clean. Its 851 traces include several failure codes with fewer than twenty positive examples. Supervised training on these annotations for failure-code prediction would lead the judge to learn framework fingerprints as shortcuts and tie its predictions to the framework mix represented in training. The performance drop of the supervised structural classifier under leave-one-framework-out evaluation in Section~\ref{sec:experiments:lofo} illustrates this dependence. We therefore keep the judge's parameters fixed and express topology knowledge through aggregate failure frequencies and structural signatures in the prompt.
\CNtranslation{4.2 备注一}{%
\textbf{备注一是零样本设计。} 零样本设计支持在 MAST-clean 标签覆盖不均衡的条件下进行诊断。其八百五十一条轨迹中，有多个失败码的正例少于二十条。若使用这些标注监督训练失败码预测器，判官会将框架指纹学习为捷径，并使预测依赖训练数据所代表的框架组合。第~\ref{sec:experiments:lofo} 节中监督结构分类器在留一框架评测下的性能下降说明了这种依赖。因此，我们保持判官参数固定，并通过提示中的聚合失败频率和结构签名表达拓扑知识。}

\CompressionEdit{%
\textbf{Remark 2. Role of topology context.} Topology context combines historical failure frequencies with an interpretation of structural failure patterns. The prior supplies corpus-level frequencies estimated once on the disjoint exploration slice and then frozen. Each failure signature uses only topology-level roles and interaction patterns to describe how structural failures arise. Examples include role dialectic loops in Layered systems, verifier bottlenecks in Centralized systems, unclaimed verification duties in Shared Message Pool systems, and peer critique drift in Decentralized systems. The prompt presents all fourteen rates as soft priors, and trace evidence guides the judge's independent decision for each code.
}{%
\textbf{Remark 2. Role of topology context.} Failure signatures complement historical frequencies with explanations based only on topology-level roles and interactions. Examples include role dialectic loops in Layered systems, verifier bottlenecks in Centralized systems, unclaimed verification duties in Shared Message Pool systems, and peer critique drift in Decentralized systems.}
\CNtranslation{4.2 备注二}{%
\CompressionEdit{\textbf{备注二是拓扑背景的作用。} 拓扑背景将历史失败频率与结构性失败模式的解释相结合。先验提供在互不重叠的探索切片上一次性估计并冻结的语料级频率。每个失败签名仅使用拓扑级角色和交互模式来描述结构性失败如何产生。例如，分层系统中的角色辩证循环、中心化系统中的校验者瓶颈、共享消息池系统中无人承担的验证职责，以及去中心化系统中的同伴批评漂移。提示将全部十四项比率作为软先验呈现，轨迹证据则指导判官对每个失败码独立作出判断。}{%
\textbf{备注二是拓扑背景的作用。} 失败签名仅依据拓扑层面的角色与交互提供解释，作为历史频率的补充。例如，分层系统中的角色辩证循环、中心化系统中的校验者瓶颈、共享消息池系统中无人承担的验证职责，以及去中心化系统中的同伴批评漂移。}}

\providecommand{\ExperimentsSection}{sections/05_experiments}

\section{Experiments}
\label{sec:experiments}  This section evaluates how the topology signal established in Section~\ref{sec:empirical} supports failure diagnosis through the two-stage pipeline of Section~\ref{sec:pipeline}.     We first describe the evaluation corpora and comparison protocol in Sections~\ref{sec:experiments:dataset} and~\ref{sec:experiments:setup}, then assess topology recovery in Section~\ref{sec:experiments:tse}.   The diagnostic gain from topology conditioning is measured in Section~\ref{sec:experiments:indist}, and cross-framework performance is examined in Section~\ref{sec:experiments:lofo}.   Finally, Section~\ref{sec:experiments:e2ecost} evaluates the complete pipeline with predicted topology and analyses its deployment cost.

\subsection{ \allowbreak{}\textsf{MAScope-Bench} Construction and Audits}
\label{sec:experiments:dataset}     \textsf{MAScope-Bench}\footnote{Code and reproducibility artifacts are available at \url{https://anonymous.4open.science/r/mascope-anonymous-artifact-6F61}. The datasets are available at \url{https://huggingface.co/datasets/mascope/mascope-bench}.} combines the three corpora introduced in Section~\ref{sec:empirical:corpora} and uses a shared schema for all \num{2220} traces. \textsf{MAST-clean} supports the main evaluation, \textsf{Inject-Mix} broadens coverage, while \textsf{LG-Ctrl} isolates topology within one framework. Their composition is summarised in Table~\ref{tab:fingerprint}.

    \textsf{MAST-clean} is selected under a uniform protocol whose pre-registered admission rule requires agreement among framework papers, the Guo survey mapping, and independent per-trace LLM voting.   The admitted traces are deduplicated by task, framework, and trajectory hash, with a stratified subset independently relabelled.  These traces retain the original MAST-14 failure annotations and provide reliable topology labels for the main diagnostic comparison.

   \textsf{Inject-Mix} uses three types of executable orchestrators spanning Layered, Centralized, and Decentralized topologies.  Twelve deterministic code-level injectors cover four failure categories by modifying shared state through runtime hooks with fixed MAST-code mappings.  The injected traces are then filtered by their targeted health scores, retaining only those with scores below the pre-set threshold of $0.55$.  The additional orchestrators and Decentralized traces extend coverage beyond \textsf{MAST-clean} to test whether topology conditioning remains useful in these settings. Because its failure labels are injection-derived, diagnostic scores are compared only within \textsf{Inject-Mix}.

   \textsf{LG-Ctrl} generates four topology variants within one LangGraph codebase, keeping the state schema, log format, and agent identifiers fixed while varying only graph edges and iteration patterns.   Its design grid samples eight LLM models, seven injectors, and a \texttt{normal} baseline near-uniformly, yielding \num{102} traces per topology. Since \textsf{MAST-clean} pairs each framework with only one topology, \textsf{LG-Ctrl} tests recovery with framework identity held fixed.

\subsection{ \allowbreak{}Comparison Methods and Evaluation Setup}
\label{sec:experiments:setup}   We compare \TCJudge with the five baselines in Table~\ref{tab:baselines}, namely B1 \textsc{Opus-Direct}, B2 $M_0$, B3 \textsc{MAST-Judge}, B4 \textsc{Flat-GBM}, and B5 \textsc{Neural-Trace}. B1, B2, and B3 provide LLM references, while B4 and B5 learn from execution graphs and trace text. Who\&When~\cite{zhang2025whowhen}, A2P~\cite{west2025a2p}, and LumiMAS~\cite{lumimas2026} are discussed qualitatively in Section~\ref{sec:related:diagnosis} because they output failure locations or free-text causes rather than MAST-14 vectors.

\begin{table}[!b]
  \centering
  \small
  \setlength{\tabcolsep}{4pt}
  \renewcommand{\arraystretch}{1.05}
  \caption{Baselines used in the main and cross-framework comparisons. All inputs are constructed from the same raw traces.}
  \label{tab:baselines}
  \begin{tabular}{@{}>{\raggedright\arraybackslash}p{0.32\columnwidth}>{\leavevmode}p{0.60\columnwidth}@{}}
    \toprule
    Baseline & Construction and comparison role \\
    \midrule
    B1 \textsc{Opus-Direct}
      & Complete trace, zero-shot model-capacity reference \\
    B2 $M_0$
      & Trace with MAST-14 schema, zero-shot matched control \\
    B3 \textsc{MAST-Judge}~\cite{cemri2025mast}
      & Trace with MAST-14 schema, original MAST judge prompt \\
    B4 \textsc{Flat-GBM}~\cite{ke2017lightgbm}
      & Seventeen topology and shallow trace features, five-fold learner \\
    B5 \mbox{\textsc{Neural-Trace}~\cite{xiao2024bge}}
      & Frozen BGE trace embedding, five-fold textual learner \\
    \bottomrule
  \end{tabular}

\end{table}

\begin{figure}[!b]
  \centering
  \includegraphics[width=\linewidth]{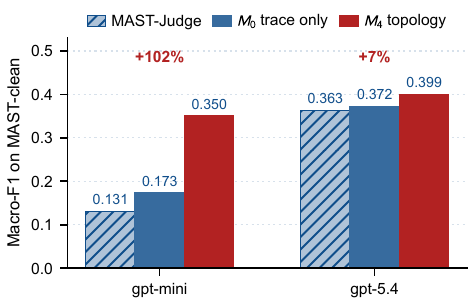}
  \caption{ Controlled diagnostic gains from $M_0$ to $M_4$ on \textsf{MAST-clean}.  Each comparison fixes the traces, prompt scaffolding, output space, and  model. Both  models improve, with gpt-mini gaining $102\%$.}
  \label{fig:m0m4-lift}
  \Description{A grouped bar chart comparing MAST-Judge, the topology-free M-zero judge, and the topology-conditioned M-four judge on gpt-mini and gpt-5.4. M-four raises Macro-F1 from 0.173 to 0.350 for gpt-mini and from 0.372 to 0.399 for gpt-5.4.}
\end{figure}


 The comparison uses the \num{851} \textsf{MAST-clean} traces, which pair original MAST-14 failure annotations with audited, stable topology labels. Macro-F1 over fourteen codes gives tail modes weight under label imbalance. The zero-shot judges \TCJudge, $M_0$, and MAST-Judge require no fitting. \textsc{Flat-GBM} and \textsc{Neural-Trace} use deterministic five-fold \texttt{GroupKFold} by \texttt{task\_id}, keeping each task in one fold. Cross-framework evaluation reports leave-one-framework-out results, denoted LOFO, for each held-out framework. \textsf{LG-Ctrl} and \textsf{Inject-Mix} provide separate structural and diagnostic validation, respectively.

   Each baseline derives its inputs from the same raw traces and predicts the same MAST-14 label vector. B1 sends the complete trace to an Opus-tier model in one call. B2, $M_0$, matches $M_4$ in traces, prompt scaffolding, output format, and model, but receives no topology context. This paired comparison isolates gains from topology conditioning. B3 reproduces the original MAST prompt~\cite{cemri2025mast}, adapts its output to MAST-14, and uses the same model as $M_0$. B4 fits LightGBM~\cite{ke2017lightgbm} to eight topology descriptors and nine shallow trace statistics, including trace length, line count, and six keyword indicators. B5 embeds concatenated traces with frozen BGE-small-en-v1.5~\cite{xiao2024bge} and trains a two-layer MLP on MAST-14 labels. Cross-method scores measure overall diagnostic utility because representations, training, and evaluation settings differ.

   The internal comparison uses five nested configurations. $M_0$ receives no topology context, $M_1$ adds the topology tag, $M_2$ adds the prior instead, $M_3$ combines the tag and prior, and $M_4$ further adds the failure signature.
   
   We conduct the paired $M_0$--$M_4$ comparison separately with gpt-mini and gpt-5.4 to examine whether the gain differs between the two judge models.
   The combined gain from $M_0$ to $M_4$ for each model is reported in Table~\ref{tab:m0m4-indist}.
  We define \textsc{Oracle-topology} as the matched $M_4$ judge supplied with ground-truth topology.
   The same $M_4$ judge is evaluated with ground-truth topology and with topology predicted by \TSE to assess the effect of topology recovery errors on diagnosis.

\subsection{ \allowbreak{}Topology Recovery}
\label{sec:experiments:tse}  Before evaluating topology-conditioned diagnosis, we first verify whether \TSE can recover the structural input required by the judge.
Recovery is evaluated in two complementary settings summarised in Table~\ref{tab:tse}, with \textsf{MAST-clean} measuring recovery on natural traces from three frameworks against their declared orchestration graphs and \textsf{LG-Ctrl} providing the framework-controlled setting described in Section~\ref{sec:experiments:dataset}. The \textsf{MAST-clean} evaluation uses \TSE with gpt-5.4 on all traces. On \textsf{LG-Ctrl}, the released full-set \TSE evaluation uses Sonnet 4.6 and is supplemented by a matched \TSE replication with gpt-5.4 on a balanced \num{40}-trace subset.

 On \textsf{MAST-clean},  \TSE correctly recovers \num{798} of \num{851} topologies for $93.8\%$ accuracy and abstains on four traces. Errors concentrate on ChatDev, where a Layered chain is occasionally classified as Centralized when the first agent takes a stronger coordinating role than the framework specification suggests.
    On \textsf{LG-Ctrl}, the full-set \TSE evaluation recovers \num{312} of \num{408} topologies for $76.5\%$ accuracy, remaining above the $25\%$ balanced chance level after framework cues are controlled, with Shared Message Pool the most difficult controlled class at $56.9\%$.
   A matched \TSE replication recovers \num{31} of \num{40} traces for $77.5\%$ accuracy, supporting the conclusion that topology remains recoverable without framework identity, although the task becomes harder.

\begin{table}[!htbp]
  \centering
  \small
  \setlength{\tabcolsep}{2pt}
  \renewcommand{\arraystretch}{1.30}
  \caption{ Topology recovery by framework and controlled topology.  C, W, and A denote correct, wrong, and abstained predictions.}
  \label{tab:tse}
  \begin{tabular*}{\columnwidth}{@{\extracolsep{\fill}}lrrr@{}}
    \toprule
    Evaluation slice & $N$ & C, W, A & Accuracy \\
    \midrule
    \multicolumn{4}{@{}l}{\emph{\textsf{MAST-clean} by framework}} \\
    \quad ChatDev   & 256 & 211, 44, 1 & 82.4\% \\
    \quad Magentic  & 165 & 160, 5, 0  & 97.0\% \\
    \quad MetaGPT   & 430 & 427, 0, 3  & 99.3\% \\
    \textbf{\textsf{MAST-clean} overall}
      & \textbf{851} & \textbf{798, 49, 4} & \textbf{93.8\%} \\
    \midrule
    \multicolumn{4}{@{}l}{\emph{\textsf{LG-Ctrl} by construction topology}} \\
    \quad Layered & 102 & 89, 1, 12 & 87.3\% \\
    \quad Centralized & 102 & 91, 11, 0 & 89.2\% \\
    \quad Decentralized & 102 & 74, 12, 16 & 72.5\% \\
    \quad Shared Message Pool & 102 & 58, 26, 18 & 56.9\% \\
    \textbf{\textsf{LG-Ctrl} overall}
       & \textbf{408} & \textbf{312, 50, 46} & \textbf{76.5\%} \\
    \bottomrule
  \end{tabular*}

\end{table}

\subsection{ \allowbreak{}Zero-Shot Diagnostic Performance}
\label{sec:experiments:indist}
The evaluation of \TCJudge examines whether topology context improves failure diagnosis. On \textsf{MAST-clean}, ground-truth topology context improves both judge models, with a larger gain for gpt-mini, as shown in Figure~\ref{fig:m0m4-lift}. From $M_0$ to $M_4$, gpt-mini's Macro-F1 rises from $0.173$ to $0.350$, a relative gain of $102\%$, while gpt-5.4 rises from $0.372$ to $0.399$. The gpt-mini result reaches $94\%$ of the trace-only gpt-5.4 score. Under the matched protocol in Section~\ref{sec:experiments:setup}, these differences isolate topology conditioning. $M_4$ also exceeds MAST-Judge on the same traces and models by $0.219$ and $0.036$, respectively, as reported in Table~\ref{tab:m0m4-indist}. These oracle results provide the reference for Section~\ref{sec:experiments:e2ecost}. Supervised learners' in-distribution and cross-framework results are reported together in Section~\ref{sec:experiments:lofo}.

\begin{figure}[!b]
  \centering
  \includegraphics[width=0.90\linewidth]{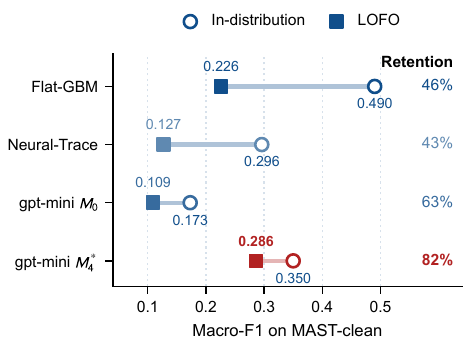}
  \caption{ Cross-framework scores on \textsf{MAST-clean}.  Flat-GBM and Neural-Trace use strict LOFO retraining.  $M_0$ has no fitted component.  $M_4^{*}$ uses a full-corpus topology prior and is a descriptive framework slice rather than a strict LOFO result.  Right labels report retention.}
  \label{fig:lofo}
  \Description{A dumbbell chart comparing in-distribution and leave-one-framework-out Macro-F1. Flat-GBM falls from 0.490 to 0.226 and Neural-Trace from 0.296 to 0.127, while descriptive zero-shot slices for M-zero and M-four retain 63 percent and 82 percent respectively.}

\end{figure}

\subsection{ \allowbreak{}Cross-Framework Generalisation}
\label{sec:experiments:lofo}  Cross-framework evaluation tests whether supervised diagnostic models transfer to deployments on an unseen framework. Most public LLM-MAS traces come from a small set of frameworks and carry distinctive signatures, such as LangGraph node names, AutoGen~\cite{wu2024autogen} tool-call ordering, and ChatDev~\cite{qian2024chatdev} role prefixes. These framework fingerprints provide distribution-specific cues to a classifier. Building on existing taxonomy and attribution benchmarks \cite{cemri2025mast,zhang2025whowhen,west2025a2p}, we use LOFO evaluation to measure robustness under this change in distribution.

  Each LOFO fold evaluates one complete framework family and trains the supervised baseline on the other two families, with all traces from the held-out framework excluded from fitting. In \textsf{MAST-clean}, each framework is paired with one topology, so the protocol introduces a joint framework and topology shift rather than isolating either factor. The mean across the ChatDev, Magentic, and MetaGPT folds and retention, defined as LOFO Macro-F1 divided by in-distribution Macro-F1, are reported in Figure~\ref{fig:lofo}. The learned baselines show substantial score reductions: Flat-GBM falls from $0.490$ to $0.226$, a relative decrease of $54\%$, while Neural-Trace falls from $0.296$ to $0.127$, retaining $43\%$ of its in-distribution Macro-F1. Flat-GBM's released confusion matrix shows that predictions on the held-out framework concentrate on the two failure modes most frequent in the retained folds. Thus, when the framework and topology change together, the strong in-distribution fit of both learned representations does not transfer reliably. The two zero-shot rows provide descriptive references and are excluded from the strict LOFO conclusion.

 The diagnostic comparison on \textsf{Inject-Mix} covers framework and topology settings not covered by \textsf{MAST-clean}.    Compared with the matched $M_0$, $M_4$ raises Macro-F1 on gpt-mini from $0.304$ to $0.321$ and on gpt-5.4 from $0.320$ to $0.356$. These results show that the diagnostic benefit of topology context extends to the additional framework and topology settings in \textsf{Inject-Mix}.

\subsection{ \allowbreak{}End-to-End Diagnosis and Deployment Cost}
\label{sec:experiments:e2ecost}  The end-to-end experiment assesses diagnostic performance when the topology predicted by \TSE is passed to \TCJudge. On \textsf{MAST-clean}, the gpt-mini pipeline reaches Macro-F1 $0.346$ compared with $0.350$ when the same judge receives ground-truth topology, giving $\Delta_{\text{oracle}} = -0.004$.  This small gap indicates that topology-recovery error has little effect on downstream diagnosis at the current evaluation scale.      

  Deployment cost is determined by the different call frequencies of structural extraction and runtime diagnosis. The cost comparison in Figure~\ref{fig:deployment-cost} uses measured gpt-5.4 diagnosis cost and the gpt-mini estimate based on normalized token counts. Under a fixed orchestration structure, deployment cost comprises one \TSE extraction per MAS and one gpt-mini \TCJudge call per trace. Reusing topology amortises extraction costs, so the cost advantage over repeated gpt-5.4 diagnosis grows with deployment volume.

\begin{figure}[!b]
  \centering
  \includegraphics[width=\linewidth]{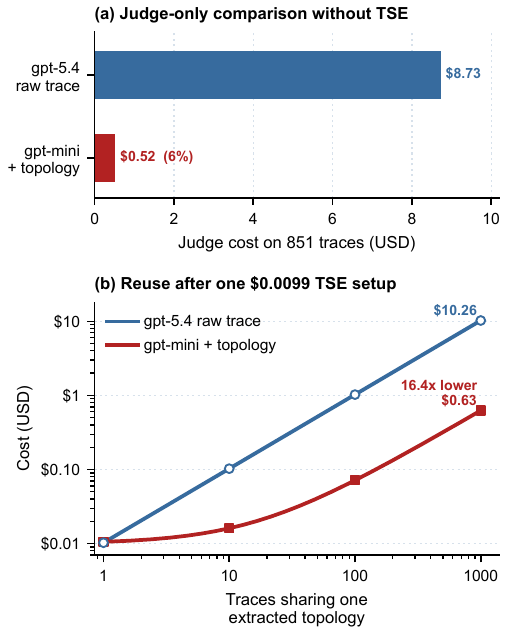}
  \caption{Cost comparison in USD. The upper panel reports measured gpt-5.4 cost and the normalized-token gpt-mini estimate on \num{851} traces without \TSE. The lower panel adds one \TSE call and shows deployment cost when the extracted topology is reused.}
  \label{fig:deployment-cost}
  \Description{Two cost charts. The upper bar chart compares 8.73 dollars for gpt-5.4 raw-trace diagnosis with 0.52 dollars for gpt-mini using topology on 851 traces. The lower log-scale line chart shows that reusing one extracted topology costs 0.63 dollars for one thousand gpt-mini diagnoses versus 10.26 dollars for gpt-5.4.}

\end{figure}


\section{Analysis}
\label{sec:analysis}

This section discusses why topology provides diagnostic information,
why it helps gpt-mini more than gpt-5.4, and how this benefit lowers
deployment cost.

\CNtranslation{6 章首}{%
本章讨论拓扑为何能够提供诊断信息，
为什么它对 gpt-mini 的帮助大于 gpt-5.4，
以及这一收益如何降低部署成本。
}

\paragraph{Q1. Why is topology informative beyond framework-specific
trace patterns?}
Communication topology plays a role analogous to organizational
structure in human teamwork. Organization theory views structure as a
means of coordinating interdependent tasks and processing information,
while coordination theory defines coordination through the management
of dependencies among activities
~\cite{galbraith1974organization,malone1994coordination}.
MAS research similarly describes an agent organization through the
roles and relationships that guide interaction and data flow
~\cite{horling2004organizational}.
Consider a project that omits a requirement. Diagnosing the problem
requires knowing who collected the requirement, who should have passed
it on, and who was responsible for checking it. Topology gives the
judge a reference for assessing messages against the collaboration's
information flows and responsibilities. This reference directs
attention to successive handoffs in Layered workflows and to the
hub's aggregation and verification in Centralized coordination.
The judge can then use trace evidence to check whether agents
fulfilled these responsibilities and determine the corresponding
failure types.

\CNtranslation{6 Q1}{%
\paragraph{Q1. 为什么拓扑能够在框架特有的轨迹模式之外提供诊断信息？}
通信拓扑在多智能体协作中的作用，类似于组织结构在人类团队中的作用。
组织理论将结构视为协调相互依赖的任务和处理信息的方式，
协调理论则通过管理活动之间的依赖来定义协调
~\cite{galbraith1974organization,malone1994coordination}。
MAS 研究也通过指导交互与数据流的角色和关系来描述智能体组织
~\cite{horling2004organizational}。
以项目遗漏一项要求为例，理解这一问题需要知道谁收集了要求、
谁应当转交，以及谁负责检查。
拓扑为判官提供了参照，使其能够依据协作中的信息流与职责评估消息。
这一参照引导判官关注分层流程中的连续交接，
以及中心化协作中中心节点的汇总与验证。
随后，判官能够利用轨迹证据检查智能体是否履行这些职责，
并确定相应的失败类型。
}

\paragraph{Q2. Why does topology help gpt-mini more than gpt-5.4?}
Without topology, the judge must first infer the collaboration structure
before deciding whether a message violates an agent's role or interrupts
an information path. Topology conditioning supplies this intermediate
structure directly. Its label identifies the coordination pattern,
while its prior and failure signature direct attention to vulnerable
responsibilities. The judge can therefore focus on verifying evidence
in the trace instead of reconstructing the collaboration scheme.
gpt-5.4 can infer more structural context from the raw trace, whereas
gpt-mini benefits more when that context is explicit.
Section~\ref{sec:experiments:indist} reports the paired diagnostic
results for both judge models.
\CNtranslation{6 Q2}{%
\paragraph{Q2. 为什么拓扑对 gpt-mini 的帮助大于 gpt-5.4？}
在没有拓扑时，判官必须先推断协作结构，
才能判断一条消息是否违背智能体职责，或是否中断信息传递路径。
拓扑条件化直接提供这一中间结构。
拓扑标签标明协作模式，先验与失败签名则引导判官关注容易出现问题的职责。
因此，判官可以集中验证轨迹中的证据，而不必先重建协作方式。
gpt-5.4 能够从原始轨迹中推断更多结构上下文，
而显式给出这些上下文对 gpt-mini 的帮助更大。
第~\ref{sec:experiments:indist} 节报告了两种判官模型的配对诊断结果。
}

\paragraph{Q3. Why does this design reduce deployment cost?}
Structural extraction and failure judgement occur at different
frequencies. A deployed MAS usually retains one orchestration graph
across many executions, so \TSE runs once and its topology is reused.
Each new trace then requires only one gpt-mini \TCJudge call. The total
cost for $n$ traces is therefore one extraction plus $n$ lower-cost
judge calls, rather than $n$ gpt-5.4 calls.
Section~\ref{sec:experiments:e2ecost} presents the deployment-cost
comparison for this reuse setting. As more traces share the fixed
extraction cost, the average cost per trace decreases.
\CNtranslation{6 Q3}{%
\paragraph{Q3. 为什么这一设计能够降低部署成本？}
结构抽取与失败判断的发生频率不同。
部署后的 MAS 通常会在多次执行中保持同一个编排图，
因此 \TSE 只需运行一次，其拓扑可以被后续执行复用。
每条新轨迹只需要调用一次 gpt-mini \TCJudge。
所以，处理 $n$ 条轨迹的总成本是一次结构抽取加
$n$ 次较低成本的判官调用，而不是 $n$ 次 gpt-5.4 调用。
第~\ref{sec:experiments:e2ecost} 节给出了这一复用设置下的部署成本比较。
随着更多轨迹分摊固定的抽取成本，每条轨迹的平均成本随之下降。
}

\section{Conclusion}
\label{sec:conclusion}

In this paper, we establish an empirical association between
communication topology and failure patterns in LLM-based multi-agent
systems. Building on this finding, we propose the two-stage diagnosis
framework \sysname{}, which recovers topology from heterogeneous
traces and combines it with empirical failure priors to guide failure
judgement. In addition, we construct and release \textsf{MAScope-Bench},
which combines natural, injected, and framework-controlled traces for
evaluating topology recovery and failure diagnosis.
Experiments show that topology conditioning improves diagnostic
performance, allowing a lower-cost judge to approach the diagnostic
performance of a stronger trace-only model, while structural reuse
reduces deployment cost.

\CNtranslation{7 研究总结}{%
本文通过实证分析建立了基于 LLM 的多智能体系统中通信拓扑与失败模式之间的关联。
基于这一发现，我们提出两阶段诊断框架 \sysname{}，
从异构轨迹中恢复拓扑，并将其与经验失败先验相结合，以指导失败判断。
此外，我们构建并发布 \textsf{MAScope-Bench}，
整合自然轨迹、注入轨迹和框架受控轨迹，用于评估拓扑恢复与失败诊断。
实验表明，拓扑条件化能够改善诊断表现，
使成本较低的判官接近更强的仅使用轨迹的模型的诊断表现，
同时，结构复用降低了部署成本。
}

\BilingualDocumentEnd

\bibliographystyle{ACM-Reference-Format}
\bibliography{references}

\end{document}